\documentclass[12pt]{article}
\usepackage{amsmath,amsfonts,amssymb}
\makeatletter
\def\numberbysection{\@addtoreset{equation}{section}
    \def\theequation{\thesection.\arabic{equation}}}
\makeatother

\numberbysection

\newcommand{\be}{\begin{eqnarray}}
\newcommand{\ee}{\end{eqnarray}}
\newcommand{\non}{\nonumber}

\begin{document}

\begin{titlepage}
\rightline{UMTG--264}
\rightline{DCPT--09/65}
\vspace{.5in}
\begin{center}

\LARGE On the absence of reflection in $AdS_{4}/CFT_{3}$\\
\vspace{1in}
\large Changrim Ahn \footnote{
       Department of Physics, Ewha Womans University,
       Seoul 120-750, South Korea; ahn@ewha.ac.kr},
       Patrick Dorey \footnote{
       Department of Mathematical Sciences,
       Durham University, Durham DH1 3LE, UK; p.e.dorey@durham.ac.uk} and
       Rafael I. Nepomechie \footnote{
       Physics Department, P.O. Box 248046, University of Miami,
       Coral Gables, FL 33124 USA; nepomechie@physics.miami.edu}\\

\end{center}

\vspace{.5in}

\begin{abstract}
A noteworthy feature of the $S$-matrix which has been proposed for
$AdS_{4}/CFT_{3}$ is that the scattering of an $A$-particle
(``soliton'') with a $B$-particle (``antisoliton'') is reflectionless.
We argue, following Zamolodchikov, that the absence of reflection is a
result of the existence of certain local conserved charges which act
differently on the two types of particles.
\end{abstract}

\end{titlepage}

\setcounter{footnote}{0}

\section{Introduction}\label{sec:intro}

The $AdS_{4}/CFT_{3}$ correspondence \cite{ABJM} relates type IIA
superstring theory on $AdS_4\times CP^3$ and planar three-dimensional
${\cal N}=6$ superconformal Chern-Simons gauge theory.  Evidence for
integrability has been found for both the string theory 
\cite{AF2}-\cite{KVV} and the gauge theory \cite{MZ}-\cite{MSS}.  
Based on the symmetries and the spectrum of
elementary excitations \cite{MZ, NT, GGY, GHO}, an all-loop $S$-matrix has
been proposed \cite{AN1} (paralleling the one \cite{St}-\cite{AFZ} for
$AdS_{5}/CFT_{4}$) which leads to the all-loop asymptotic Bethe ansatz
equations proposed in \cite{GV2}.

The elementary excitations consist of so-called $A$-particles and
$B$-particles.  Since these particles are related by $CP$ symmetry,
they can be regarded as ``solitons'' and ``antisolitons,''
respectively \cite{Zar}.  A noteworthy feature of the proposed
$AdS_{4}/CFT_{3}$ $S$-matrix is that the scattering of an $A$-particle
with a $B$-particle is reflectionless.  This property has been
explicitly verified both at weak coupling \cite{AN2} and at strong
coupling \cite{Zar}.  However, an explanation for this property has
been missing.

A possible clue comes from the observation that various integrable
relativistic (1+1)-dimensional quantum field theories share this
feature of reflectionless scattering.  One example is the thermal
perturbation of the 3-state Potts model \cite{KS}, with an S-matrix
which is related to the
$A_{2}$ affine Toda field theory (ATFT).  Zamolodchikov \cite{Zam}
showed that the absence of reflection in this model is a result of the
existence of higher spin
(greater than 1) local integrals of motion which act differently on
the particle and on the antiparticle.  Similar examples are
provided by the thermal perturbation of the tricritical 3-state Potts
model, which is related to the $E_{6}$ ATFT \cite{FZ}; 
and the $D_{4}$ ATFT \cite{BCDS}.

In this note, we propose an analogous explanation for the absence of 
reflection in $AdS_{4}/CFT_{3}$. In Section \ref{sec:charges}, we 
identify certain local conserved charges, and argue that they imply 
the reflectionless property. In Section \ref{sec:curve} we reach the 
same conclusion from consideration of the algebraic curve. Finally, 
in Section \ref{sec:discussion} we briefly discuss our results.

\section{Local charges and the reflectionless property}\label{sec:charges}

The two-loop dilatation operator of ${\cal N}=6$ superconformal
Chern-Simons gauge theory has been studied in \cite{MZ}-\cite{MSS}.
For the $su(4)$ sector \cite{MZ, BR}, there are two associated
commuting transfer matrices $\tau(u)\,, \bar\tau(u)$.  That is, these
transfer matrices obey
\be
\left[ \tau(u)\,, \tau(v) \right] = 0\,, \qquad
\left[ \tau(u)\,, \bar\tau(v) \right] = 0\,, \qquad
\left[ \bar\tau(u)\,, \bar\tau(v) \right] = 0\,, 
\ee
for arbitrary values of the spectral parameters $u$ and $v$. Hence, 
there are two sets of local charges,
\be
Q_{n} = \frac{d^{n-1}}{du^{n-1}}\ln \tau(u) \Big\vert_{u=0}\,, \qquad
\bar Q_{n} = \frac{d^{n-1}}{du^{n-1}}\ln \bar\tau(u) \Big\vert_{u=0} 
\,, \quad n = 1, 2, \ldots
\ee
which are mutually commuting,
\be
\left[ Q_{n}\,, Q_{m} \right] =  0\,, \qquad
\left[ Q_{n}\,, \bar Q_{m} \right] =  0\,, \qquad
\left[ \bar Q_{n}\,, \bar Q_{m} \right] =  0 \,.
\ee
The Hamiltonian (dilatation operator) is proportional to $Q_{2} + 
\bar Q_{2}$, and therefore all the charges are conserved. The 
eigenvalues of the charges are given by
\be
Q_{n} &=& \frac{i}{n-1}\sum_{j=1}^{K_{4}}
\left(\frac{1}{(u_{4,k}+\frac{i}{2})^{n-1}}
-\frac{1}{(u_{4,k}-\frac{i}{2})^{n-1}} \right) \,, \non \\
\bar Q_{n} &=& \frac{i}{n-1}\sum_{j=1}^{K_{\bar 4}}
\left(\frac{1}{(u_{\bar 4,k}+\frac{i}{2})^{n-1}}
-\frac{1}{(u_{\bar 4,k}-\frac{i}{2})^{n-1}} \right) \,,
\label{eigenvalues}
\ee
where $u_{4,k}$ and $u_{\bar 4,k}$ are the ``momentum-carrying'' Bethe
roots.  The same is true for the $osp(4|2)$ sector, as well as for the
full two-loop model \cite{Zw}.  
We shall assume that these charges lift to the all-loop asymptotic model, 
as in Section 7 of \cite{BBF},
\footnote{We are grateful to B. Zwiebel for correspondence on this
point.} to yield conserved charges with eigenvalues 
\be
Q_{n} &=& \frac{i}{n-1}\sum_{j=1}^{K_{4}}
\left(\frac{1}{(x_{4,k}^{+})^{n-1}}
-\frac{1}{(x_{4,k}^{-})^{n-1}} \right) \,, \non \\
\bar Q_{n} &=& \frac{i}{n-1}\sum_{j=1}^{K_{\bar 4}}
\left(\frac{1}{(x_{\bar 4,k}^{+})^{n-1}}
-\frac{1}{(x_{\bar 4,k}^{-})^{n-1}} \right) \,,
\label{assumption}
\ee
where 
\be
x+\frac{1}{x} &=& \frac{u}{h(\lambda)} \,, \non \\
x^{\pm} +\frac{1}{x^{\pm}} &=& \frac{u\pm \frac{i}{2}}{h(\lambda)}\,.
\ee
The still-unknown function $h$ of the 't Hooft parameter $\lambda$
must satisfy $h(\lambda) \simeq \lambda$ for $\lambda \ll 1$, and
$h(\lambda) \simeq \sqrt{\lambda/2}$ for $\lambda \gg 1$.
In proposing the existence of a set of charges with eigenvalues 
(\ref{assumption}) in the full theory, we are making a
stronger claim for the conserved charges than~\cite{GV2}, where 
it was claimed that the spectrum of all conserved charges is given by
a set of numbers $\{ {\cal Q}_n\}$, where
${\cal Q}_n=Q_n+\bar Q_n$ in our notations. 
Purely at the level of the spectrum,
one might argue that the value of $Q_n$ for a state with
$K_4$ $u_4$ roots and $K_{\bar 4}$ $u_{\bar 4}$ roots could be 
replicated as the value of ${\cal Q}_n$ for
a state with the same number of $u_{4}$ roots and 
no $u_{\bar 4}$ roots, but this is not the case: the two types of roots
interact in the full set of Bethe ansatz equations, and hence affect
each other's positions. Though somewhat subtle, this difference from
\cite{GV2} is crucial to our understanding of the reflectionless
property, because it allows us to distinguish between the $A$ and $B$
particles.

We now introduce (as in \cite{AN1}) Zamolodchikov-Faddeev operators
$A_{i}^{\dagger}(p)$ and $B_{i}^{\dagger}(p)$ ($i = 1, \ldots, 4$)
corresponding to $A$-particles and $B$-particles, respectively.  These
particles are associated with the two types of momentum-carrying Bethe
roots $u_{4}$ and $u_{\bar 4}$, respectively. (The momenta are given 
by $e^{i p}=\frac{x_{4}^{+}}{x_{4}^{-}}$ and 
$e^{i p}=\frac{x_{\bar 4}^{+}}{x_{\bar 4}^{-}}$ , respectively.)
It follows from (\ref{assumption}) that
\be
Q_{n}\, A_{i}^{\dagger}(p)\, |0\rangle &=& q_{n}(p)\, 
A_{i}^{\dagger}(p)\, |0\rangle\,, \qquad 
q_{n}(p) = \frac{i}{n-1}\left( \frac{1}{(x_{4}^{+})^{n-1}}
-\frac{1}{(x_{4}^{-})^{n-1}} \right)\,, \non \\
Q_{n}\, B_{i}^{\dagger}(p)\, |0\rangle &=& 0 \,, 
\label{Qaction}
\ee
and
\be 
\bar Q_{n}\, A_{i}^{\dagger}(p)\, |0\rangle &=& 0  \,, \non \\
\bar Q_{n}\, B_{i}^{\dagger}(p)\, |0\rangle &=& \bar q_{n}(p)\, 
B_{i}^{\dagger}(p)\, |0\rangle \,, \qquad 
\bar q_{n}(p) = \frac{i}{n-1}\left( \frac{1}{(x_{\bar 4}^{+})^{n-1}}
-\frac{1}{(x_{\bar 4}^{-})^{n-1}} \right)\,.  
\label{barQaction}
\ee

The scattering of an $A$-particle and a $B$-particle can in principle 
have both transmission and reflection,
\be
A_{i}^{\dagger}(p_{1})\, B_{j}^{\dagger}(p_{2}) =
S_{i\, j}^{i' j'}(p_{1}, p_{2})\,
B_{j'}^{\dagger}(p_{2})\, A_{i'}^{\dagger}(p_{1}) +
R_{i\, j}^{i' j'}(p_{1}, p_{2})\,
A_{j'}^{\dagger}(p_{2})\, B_{i'}^{\dagger}(p_{1}) \,,
\label{ABscattering}
\ee
where $S$ and $R$ are the transmission and reflection amplitudes, 
respectively. Acting on (\ref{ABscattering}) with the local charge 
$Q_{n}$, and then making use of (\ref{Qaction}),  we obtain
\be
\lefteqn{q_{n}(p_{1})\, A_{i}^{\dagger}(p_{1})\, B_{j}^{\dagger}(p_{2}) 
|0\rangle} \non \\
&& = q_{n}(p_{1})\, S_{i\, j}^{i' j'}(p_{1}, p_{2})\,
B_{j'}^{\dagger}(p_{2})\, A_{i'}^{\dagger}(p_{1}) |0\rangle +
q_{n}(p_{2})\, R_{i\, j}^{i' j'}(p_{1}, p_{2})\,
A_{j'}^{\dagger}(p_{2})\, B_{i'}^{\dagger}(p_{1}) |0\rangle \,. 
\label{ABscattering2}
\ee
Since in general  $q_{n}(p_{1})$ is nonzero and
$q_{n}(p_{1}) \ne q_{n}(p_{2})$, the two equations (\ref{ABscattering}) and 
(\ref{ABscattering2}) are not compatible unless the reflection 
amplitudes vanish,
\be
R_{i\, j}^{i' j'}(p_{1}, p_{2}) = 0 \,.
\ee
Evidently, the same result can also be obtained by instead acting on
(\ref{ABscattering})  with the local charges $\bar Q_{n}$ and
making use of (\ref{barQaction}). We conclude that the existence of 
the local charges $Q_{n}$ and $\bar Q_{n}$ which act differently on 
the $A$-particles and $B$-particles implies the absence of
reflection.\footnote{In this analysis we have neglected the constraint
of zero total
momentum arising from the cyclicity of the trace \cite{MZ}.  For the
case
of two particles, this constraint implies the restriction $p_2 = -
p_1$.
Since there are functions $q_n(p)$ which are not even, the conclusion
still holds.}

\section{Algebraic curve}\label{sec:curve}

The same conclusion can be drawn from a consideration of the algebraic
curve \cite{GV1, GV2}.\footnote{The $AdS_{5}/CFT_{4}$ algebraic
curve was formulated in \cite{KMMZ}-\cite{BKSZ}; 
see~\cite{V} for a review of the general formalism.} The $A$ and $B$
particles correspond to the two orientations of certain giant
magnon solutions of the string sigma-model \cite{AAS, HM}.  In the
language of \cite{Sh}, these correspond to certain ``small'' giant
magnons.  Following \cite{AAS} (see also \cite{GV1, LS}), we assume
that the quasi-momenta are given by\footnote{These quasi-momenta
should not be confused with the functions $q_n(p)$ used in the
previous section.}
\be
q_{1}(x) &=& \frac{\alpha x}{x^{2}-1} \,, \non \\
q_{2}(x) &=& \frac{\alpha x}{x^{2}-1} \,, \non \\
q_{3}(x) &=& \frac{\alpha x}{x^{2}-1} +  G_{4}(0) -G_{4}(\frac{1}{x})
+ G_{\bar 4}(0) -G_{\bar 4}(\frac{1}{x})
+ G_{3}(x) - G_{3}(0) + G_{3}(\frac{1}{x})\,, \non \\
q_{4}(x) &=& \frac{\alpha x}{x^{2}-1} +  G_{4}(x) + G_{\bar 4}(x) 
- G_{3}(x) + G_{3}(0) - G_{3}(\frac{1}{x})\,, \non \\
q_{5}(x) &=& G_{4}(x) - G_{4}(0) + G_{4}(\frac{1}{x})
-G_{\bar 4}(x) + G_{\bar 4}(0) - G_{\bar 4}(\frac{1}{x}) \,.
\label{qmom}
\ee
The asymptotic behavior of the quasi-momenta as $x \rightarrow \infty$
determines the various quantum numbers.

In a nutshell, our argument is that the algebraic curves corresponding to 
\begin{itemize}
    \item[(i)] a particle $A$ of momentum $p$ and a particle $B$ of 
    momentum $-p$, and
    \item[(ii)] a particle $A$ of momentum $-p$ and a particle $B$ of 
    momentum $p$
    \end{itemize}
are different, since the $q_{5}(x)$ are different. Since the 
algebraic curve is a classical invariant, there cannot exist a 
solution with initial asymptotic state (i) and final asymptotic state 
(ii). 
    
In more detail, let us begin by observing that 
a single $A$-particle with momentum $p$ corresponds to
\be
G_{4}(x) = G_{mag}(x)\,, \qquad G_{\bar 4}(x) = G_{3}(x) = 0\,,
\label{curveA}
\ee
and a single $B$-particle with momentum $p$ corresponds to
\be
G_{\bar 4}(x) = G_{mag}(x)\,, \qquad G_{4}(x) = G_{3}(x) = 0\,,
\label{curveB}
\ee
where \cite{AAS, MTT}
\be
G_{mag}(x) = -i \ln \left(\frac{x - X^{+}}{x - X^{-}} \right) \,,
\ee
and $e^{i p} = \frac{X^{+}}{X^{-}}$.  One can show that $A$ and $B$
particles with momentum $-p$ correspond to (\ref{curveA}) and
(\ref{curveB}) with $G_{mag}(x)$ replaced by $\tilde G_{mag}(x)$,
respectively, where
\be
\tilde G_{mag}(x) = -i \ln \left(\frac{x + X^{-}}{x + X^{+}} \right) 
\,.
\ee
We note that for fundamental (i.e., non-dyonic, $Q=1$) giant magnons,
$q_{5}(x)$ is of order $1/g$, since $X^\pm = e^{\pm i p/2}[1 +
o(1/g)]$.  We assume that $g$ is large but finite, so that $q_{5}(x)$
is nonzero.

Let us consider an initial two-particle configuration consisting of an
$A$-particle with momentum $p$, and a $B$-particle with momentum $-p$.
As in the ansatz of \cite{LS} for the more-complicated case of
multi-magnon states in finite volume, we suppose that the 
quasi-momenta  corresponding to this initial state are simply the sums
of the quasi-momenta for the two constituent particles.
In particular, this means that $q_{5}(x)$ is
\be
q_{5}(x)\Big\vert_{initial} = 
G_{mag}(x) - G_{mag}(0) + G_{mag}(\frac{1}{x})
-\tilde G_{mag}(x) + \tilde G_{mag}(0) - \tilde G_{mag}(\frac{1}{x}) 
\,.
\ee

Now consider the possible states after the collision has occurred.
The ``transmitted'' configuration again consists of an $A$-particle
with momentum $p$ and a $B$-particle with momentum $-p$, and so
\be
q_{5}(x)\Big\vert_{transmitted} = 
q_{5}(x)\Big\vert_{initial} 
\,.
\ee
The ``reflected'' configuration consists instead of an
$A$-particle with momentum $-p$ and a $B$-particle with momentum 
$p$; and therefore
\be
q_{5}(x)\Big\vert_{reflected} &=&  
\tilde G_{mag}(x) - \tilde G_{mag}(0) + \tilde G_{mag}(\frac{1}{x})
- G_{mag}(x) +  G_{mag}(0) -  G_{mag}(\frac{1}{x}) \non \\
&=& - q_{5}(x)\Big\vert_{initial} 
\,.
\ee
Since in general $q_{5}(x)\Big\vert_{initial}$ is nonzero,
it follows that $q_{5}(x)\Big\vert_{reflected} \ne
q_{5}(x)\Big\vert_{initial}$; and therefore, reflection is not
possible.

The quasi-momenta can be expressed in terms of the scaling limit of 
the conserved charges (\ref{eigenvalues}), see e.g. \cite{GV2, BKS}.
Hence, the above computation essentially confirms that the set of classical
conserved charges is powerful enough to forbid reflection. 
Note that the quasi-momentum $q_{5}(x)$, which is of key importance in
our argument, behaves like $Q_{n} - \bar Q_{n}$ and not like $Q_{n} +
\bar Q_{n}$.  Indeed, as can be seen from Eq.  (\ref{qmom}),  $q_{5}(x)$
is given by the {\em difference} of quantities involving $G_{4}(x)$ 
and $G_{\bar 4}(x)$, which in turn depend on $X_4$ and $X_{\bar 4}$, 
respectively. In that sense, the classical argument also makes use of 
the two towers of conserved charges which were essential in the 
quantum argument of Sec. 2.

It would be valuable to extend these considerations to the full quantum
theory, by computing quantum corrections to the local charges that
appear in the expansions of the quasi-momenta about $=\pm 1$, 
but we will leave this to future work. Even remaining at the classical
level, it would be interesting to construct classical solutions corresponding
to the scattering of $A$ and $B$ particles, and verify that these
solutions do not exhibit reflection.  

It may be useful to compare the situation with the more familiar 
example of the sine-Gordon theory \cite{ZZ}. 
Classical soliton-antisoliton scattering is also
reflectionless in that case. However, unlike the $AdS_{4}/CFT_{3}$ case, 
reflection is not forbidden by conservation laws. (Hence,  in that 
sense, the absence of reflection in classical sine-Gordon theory is ``accidental.'')
Indeed, since solitons and antisolitons are only
distinguished by a spin-zero (topological) charge, there is no
obstruction to a nonzero reflection amplitude appearing in the
quantum sine-Gordon theory; and by Gell-Mann's totalitarian principle that 
in quantum theory 
`everything that is not forbidden is compulsory', the quantum
sine-Gordon theory does indeed exhibit reflection as well 
as transmission.\footnote{At a discrete set of values of the
coupling, quantum sine-Gordon is again reflectionless \cite{ZZ}.
However, at exactly those couplings, it can be argued that
the sine-Gordon model picks up an
extra conserved charge which {\em does}\/ split soliton from
antisoliton,
and so the totalitarian principle survives \cite{alyosha}.} 
(For an elementary discussion of this point, see \cite{Do}.)
The crucial difference is that in the
$AdS_4/CFT_3$ case there is a momentum-dependent conserved charge
which splits apart the $A$ and the $B$, at all values of the coupling.

\section{Discussion}\label{sec:discussion}

We have argued that the origin of the $AdS_{4}/CFT_{3}$ reflectionless
property is the existence of two commuting transfer matrices, and
therefore two sets of commuting conserved local charges.  This is 
in contrast to the $AdS_{5}/CFT_{4}$ case, for which there is only 
one commuting transfer matrix, and therefore only one set of commuting 
conserved local charges. 
The argument is a generalization of the one used in the study of 
purely elastic scattering theories \cite{Zam, FZ, BCDS}.
Our discussion of the string sigma-model side of this story has been
preliminary, and it would be of particular interest to give a
full quantum treatment. In the context of relativistic quantum field
theories, local conserved charges often hide intricate 
structures, such as the Coxeter geometry found in the purely elastic
scattering theories (see, for example, \cite{Do} and references therein). 
It would be 
interesting to see whether similar phenomena also exist in $AdS/CFT$.

\vspace{.2in}

{\em Note Added:} 
As we were about to submit this paper to the Arxiv, the paper \cite{HT}
appeared, which discusses issues related to Sec.  \ref{sec:curve}.

\section*{Acknowledgments}
We thank Michael Abbott, In\^es Aniceto and 
Benjamin Zwiebel for valuable correspondence; Simon Ross
and Charles Young for helpful comments; Konstantin 
Zarembo for reading and commenting on a preliminary draft; and
Benoit Vicedo for many careful
explanations, as well as valuable comments on a later draft. PED thanks the
University of Miami, and RN
thanks Durham University, for hospitality during the course
of this work.  This work was supported in part by KRF-2007-313-C00150
and WCU grant R32-2008-000-10130-0 (CA), by STFC rolling grant
ST/G000433/1 and a grant from the Leverhulme Trust (PED),
and by the National Science
Foundation under Grants PHY-0554821 and PHY-0854366 (RN).

\end{document}